# *Slow light or saturable absorption?*

**A C SELDEN**

Department of Physics University of Zimbabwe

MP 167 Mount Pleasant HARARE Zimbabwe

e-mail address   *acselden@science.uz.ac.zw*

**ABSTRACT**

Quantitative analysis of slow light experiments utilising coherent population oscillation (CPO) in a range of saturably absorbing media, including ruby and alexandrite, $Er^{3+}$:$Y_2SiO_5$, bacteriorhodopsin, semi-conductor quantum devices and erbium doped optical fibres, shows that the observations may be more simply interpreted as saturable absorption phenomena. A basic two-level model of a saturable absorber displays all the effects normally associated with slow light, namely phase shift and modulation gain of the transmitted signal, hole burning in the modulation frequency spectrum and power broadening of the spectral hole, arising from the finite response time of the non-linear absorption. Only where hole-burning in the optical spectrum is observed (using independent pump and probe beams), or pulse delays exceeding the limits set by saturable absorption are obtained, can reasonable confidence be placed in the observation of slow light in such experiments. Superluminal ('fast light') phenomena in media with reverse saturable absorption (RSA) may be similarly explained.

1. Introduction

It has recently been pointed out that saturable absorption can explain many of the features attributed to slow light in a non-linear absorber, such as the phase shift of an optical signal and delayed transmission of an optical pulse, without assuming hole burning via coherent population oscillations (CPO) or an associated reduction in group velocity [1]. Similar considerations apply to pulse propagation in Pb vapour [2], GaAs quantum wells [3] and InAs quantum dots [4]. We give a brief resumé of harmonic signal transmission by a saturable absorber and make quantitative comparison with the results of CPO experiments in various media [5-16] with a view to resolving this issue.

2. Theory

For weakly modulated light incident on a saturable absorber, signal distortion is negligible, the transmitted intensity acquiring a frequency dependent phase shift and modulation gain, as observed in ruby [5, 13], alexandrite [14, 15], $Er^{3+}$:$Y_2SiO_5$ [6], bacteriorhodopsin (bR) [7-9], Er-doped optical fibres [10, 16], quantum wells [11, 12] and quantum dots [4], the observed phase lag invariably being assumed to indicate a reduction in group velocity. However, both the phase shift and modulation gain can be equally well explained by the non-linear response of the saturable absorber, with its characteristic relaxation time [17], as in amplitude modulation spectroscopy [18].

For an optically thick saturable absorber with weakly modulated (m<<1) incident intensity $I_{in}(t) = I_0[1+m\cos\omega t]$, the transmitted intensity $I_{out}(t) = I_0 T_s[1+m|K(\omega)|\cos\omega t-\varphi)]$, where $T_s$ is the steady state transmission at $I_0$, given by $\ln(T_s/T_0) = \beta(1-T_s)$, $T_0$ is the initial transmission, the saturation parameter $\beta = I_0/I_s$ for saturation intensity $I_s$ and the modulation factor (amplitude and phase) [17-20]

$$K(\omega) = \frac{1+\beta(0)+i\omega\tau_s}{1+\beta(L)+i\omega\tau_s} \qquad (1)$$

where $\beta(L) = \beta(0)T_s$ for physical length L and $\tau_s$ is the relaxation time. Thus the transmitted signal has modulation gain $|K(\omega)|>1$ for $T_s<1$ and phase $\varphi = \arg K(\omega)$. Comparing $|K(\omega)|$ from eq (1) with the relative modulation attenuation $A(\omega)$ for CPO as defined in [5], we find $A(\omega) \equiv -\ln|K(\omega)|$ thus

$$A(\omega) \;=\; 0.5 \ln \frac{(1+\beta(L))^2 + (\omega\tau_s)^2}{(1+\beta(0))^2 + (\omega\tau_s)^2} \tag{2}$$

which has its minimum value $A(0) = \ln\{(1+\beta(L))/(1+\beta(0))\}$ at zero modulation frequency. For high modulation frequencies $\omega \gg 1/\tau_s$, $A(\omega) \Rightarrow 0$ and the modulation attenuation is the same as that of the pump. Given the exact agreement of the modulation amplitudes, we see that te phase shifts should also agree, since both derive from the complex gain factor $K(\omega)$. Evaluating $\varphi = \arg K(\omega)$ from eq (1)

$$\varphi \;=\; -\arctan \frac{\omega\tau_s(\beta(0) - \beta(L))}{(1+\beta(0))(1+\beta(L)) + (\omega\tau_s)^2} \tag{3}$$

which is identical with the phase shift of intensity modulated light arising from CPO induced hole burning within the absorption line [20] (App. eq (A5)). Maximum phase shift $\varphi_m$ occurs at modulation frequency $\omega_m$, where $\omega_m^2 \tau_s^2 = (1+\beta(0))(1+\beta(L))$ [17]. Hence $\tan\varphi_m = -\beta(0)/2\sqrt{(1+\beta(0))}$ for $\beta(L) \ll \beta(0)$ and $|\varphi_m| \leq 90$ deg. From eq (3), the signal delay $\tau_d = \varphi(\omega)/\omega$ in the limit of small phase shift ($\varphi \ll 1$) and high signal transmission ($1-T_s \ll 1$) is given by (App. eq (A7))

$$\tau_d = \tau_s \frac{\beta}{1+\beta} \frac{\alpha_0 L}{(1+\beta)^2 + (\omega_m \tau_s)^2} \tag{4}$$

having a Lorentzian dependence on modulation frequency, with full width at half maximum

$$\Delta\nu_{FWHM} \;=\; \frac{1+\beta}{\pi\tau_s} \tag{5}$$

i.e. linear power broadening. Thus saturable absorption theory is in precise agreement with the theory of coherent hole burning [21]. The maximum time delay in signal transmission is reached at zero modulation frequency: $\tau_d|_{\omega=0} = \tau_s\beta(1-T_s)/(1+\beta)(1+\beta T_s)$, with a limiting value $\tau_d \Rightarrow \tau_s\beta/(1+\beta) \Rightarrow \tau_s$ for $T_s \ll 1$ and $\beta \gg 1$ and the signal delay never exceeds the lifetime $\tau_s$ of the metastable state [19, 20].

For inverse (reverse) saturable absorption, where transmission decreases with increasing intensity via excited state absorption [14], the modulation amplitude is reduced, the attenuation reaching its maximum value at $\omega = 0$. The modulation phase is positive and the signal advanced with respect to the input [1]. A similar analysis for gain saturation in an optical amplifier also predicts a forward phase shift and temporal advance $\tau_a$ in signal transmission, with a maximum value at zero modulation frequency: $\tau_a|_{\omega=0} = \tau_s\beta(G_s-1)/(1+\beta)(1+\beta G_s)$, where $G_s$ is the steady state gain at intensity $I_0$ and $\ln(G_0/G_s) = \beta(G_s-1)$. Thus $\tau_a \Rightarrow \tau_s$ as $G_s \Rightarrow \infty$ i.e. the signal advance $\tau_a$ cannot exceed the lifetime $\tau_s$.

3. Comparison with experiment

In the following, the results of some well-documented CPO experiments on non-linear absorbers with relaxation times ranging from nanoseconds to seconds are compared with saturable absorption theory, using the known physical parameters of the media involved.

3.1 Ruby

Hillman et al published an extensive set of data for CPO experiments on ruby, using an amplitude modulated argon-ion laser [13]. A series of saturable absorption curves derived from eq (1), using the experimental parameters for ruby, are plotted in Fig 1, showing excellent agreement with the experimental points. In particular, the saturable absorption curves simulate 'hole burning' and 'power broadening' effects normally associated with wave interactions in saturable absorbers [22]. More recently, Bigelow et al claim to have observed slow light in ruby using a similar arrangement [5]. Fig 2 shows the frequency dependence of the relative modulation attenuation $A(\omega)$ and time delay $\tau_d(\omega)$ for intensity modulated argon laser light transmitted by a 72.5 mm long ruby crystal used in that experiment. The theoretical curves for saturable absorption achieve near perfect agreement with the experimental data over the full frequency range on adjusting the parameter $\beta$ to fit the datum $\tau_d = 612$ µs at 60 Hz for $P_0 = 0.25$W [5]. However, the plotted curves for signal delay $\tau_d$ only properly fit the experimental data if the factor ½ in eq (9) of [5] is retained i.e. the signal delay is half that predicted for saturable absorption.

3.2 Alexandrite

Alexandrite exhibits both saturable absorption and inverse saturable absorption, the latter resulting in an increase of absorption with intensity (arising from excited state absorption) and a phase advance of the transmitted signal, interpreted as evidence of fast light [14]. $Cr^{3+}$ ions occupy two sites in the alexandrite crystal, those at mirror sites experiencing excited state absorption with a relaxation time $\tau_m$ = 260 μs, those at inversion sites saturated absorption with a relatively long relaxation time $\tau_i \sim 50$ ms [15]. As a result, CPO experiments on alexandrite yield a modulation spectrum with a narrow hole (8.4 Hz HWHM) appearing within a broad anti-hole (612 Hz HWHM) [15]. Fig 3 shows the saturable absorption curves fitted to the experimental data for an argon laser pump wavelength $\lambda_p$ = 488 nm, reproducing the principal features observed and giving good quantitative agreement for the theoretical parameters in [15].

3.3 $Er^{3+}:Y_2SiO_5$

The results for CPO in $^{167}Er^{3+}:Y_2SiO_5$ [6] are shown in Fig 4, where the variation of signal delay with incident power for modulation frequencies 0, 10, 40 and 80 Hz is plotted, comparing saturable absorption theory with the experimental data points. The theoretical curves are of similar form, those for higher modulation frequency having a flatter response and lower maximum, as observed. The scatter of the experimental data for 10 and 40 Hz modulation renders the precise location of the broad maxima uncertain, but thereafter the data points follow the long tailed distribution characteristic of inhomogeneous broadening [23]. Those for 80 Hz modulation show a similar trend. The approximate 'slow light' formula for the time delay (eq (4) in [6]), with a factor ½ included, gives a good fit to the experimental data. It was found necessary to introduce the same factor in Sec 2 eq (4) for saturable absorption theory to fit the experimental data, as for ruby. Mørk et al [20] have shown that the change of group delay for broadband signals is half that expected for the double sideband modulation characteristic of CPO, which may justify retaining the ½ in the expression for the signal delay, since the 2 kHz width pump source used in the $Er^{3+}:Y_2SiO_5$ experiment was repeatedly swept over the inhomogeneous linewidth $\Gamma_{inh}$ = 1.3 GHz, making it effectively broadband [6].

3.4 Bacteriorhodopsin

CPO experiments have been conducted on bacteriorhodopsin (bR), both as bR-doped polymer film [7] and in aqueous solution [8, 9], enabling the non-linear response to be studied over timescales ranging from milliseconds to seconds. For bR film, the relaxation rate is controlled by illumination with blue light; increasing the intensity $I_b$ of $\lambda$ = 442 nm light reduces the excited state lifetime: $\tau_s'$ =

$\tau_s/(1+AI_b)$, where $A = 1.18$ mW$^{-1}$ and $\tau_s = 0.28$ sec [7]. The dependence of signal delay $\tau_d$ on $I_b$ for these parameters is shown in Fig 5, saturable absorption giving excellent agreement with experiment. The dependence of signal delay on incident power for aqueous bR is plotted in Fig 6. The fitting parameters are slightly lower than the experimental values, which may result from neglecting the effect of signal attenuation [8]. Nevertheless, it is clear that saturable absorption matches experiment here also. In a subsequent set of experiments, the observed correlation of signal delay and attenuation vs. modulation frequency (overlaid in Fig 7) was interpreted in favour of slow light [9]. However, the data points are well matched by the saturable absorption curves, which involve no such assumption, but simply reflect the close correspondence of the amplitude and phase functions (Sec 2 eq (2, 3)). The results for reverse saturable absorption are similarly well-matched (Fig 8).

3.5 <u>Semiconductor nano-structures</u>

CPO experiments utilising heavy hole (H-H) absorption in GaAs quantum wells extend the timescale to the nanosecond range. The results for transmission of RF amplitude modulated light in the 25-175 MHz frequency band by a GaAs QW waveguide are presented in Fig 9, showing the observed time delay vs modulation frequency for input powers 25, 50 and 75 mW [11]. The experimental points are fitted with Lorentzian curves, interpreted as holes in the frequency spectrum, whose width yields the relaxation time; the saturation power $P_s$ is derived from the observed power broadening, applying the formula $\Delta\nu = (1+P/P_s)/\pi\tau_s$ (Sec 2 eq (5)) Thus CPO and saturable absorption give the same results viz. $\tau_s = 5.9\pm0.3$ ns, $P_s = 123\pm13$ mW [11]. Experiments involving 16.7 GHz intensity modulated transmission in a semiconductor waveguide confirm the equivalence of the CPO and rate equation models for saturable media with controllable loss [20]. The model has also been applied to analyse signal transmission in a sequential semiconductor optical amplifier/electro-absorber architecture [24]. However, experiments on GaAs multiple quantum wells (MQW) using independent pump and probe beams have shown both saturation of the heavy hole (H-H) absorption band and the creation of a narrow hole within the background absorption, perhaps the only direct demonstration of coherent hole-burning reported thus far in a CPO experiment [12]. Even here the data can be fitted by saturable absorption theory (see Fig 10), emphasising the point that pulse delay alone is not sufficient to demonstrate slow light, without simultaneous observation of hole-burning within the absorption line.

3.6 <u>Erbium-doped optical fibres</u>

An extensive series of measurements on pulsed and intensity modulated transmission in single mode erbium-doped optical fibres reported by Melle et al [10] enables comprehensive analysis of signal transmission obtained under well-defined experimental conditions. Only results for the lower loss fibres are considered here, higher loss fibres (40 - 110 dB/m) producing anomalous results via ion-ion interactions and inhomogeneous up-conversion. Fig 11 shows excellent agreement of saturable absorption theory with the experimental data for the fractional delay $F_d$ vs. modulation frequency in the 20 dB/m fibre, with $F_d = \varphi/2\pi$ calculated from eqn (3) for the parameters $P_s$ = 0.4 mW, $\tau_s$ = 10.5 ms [10]. Plotting maximum fractional delay $F_m = \varphi_m/2\pi$ with the same parameters provides a good fit to the experimental data for $F_m$ vs. input power $P_0$ (Fig 12). The near-linear dependence of optimum modulation frequency $f_{opt}$ on input power predicted by saturable absorption theory precisely matches the experimental data (Fig 13). These examples typify the excellent results obtained for passive transmission in erbium-doped optical fibres. CPO experiments on erbium doped fibre amplifiers show a transition from signal delay to advance when the gain exceeds the absorption loss at a given pump power and modulation frequency. This too can be explained by optical saturation theory [16, 25]. The demonstration of controllable phase shifts in Er-fibre using independent pump and probe beams with widely separated wavelengths provides further support for the saturable absorption model [26].

4. <u>Discussion</u>

One of the problems with slow light experiments based on coherent population oscillations (CPO) in saturable absorbers is that they do not offer any direct evidence for group velocity reduction, as inferred from the phase shift of the transmitted signal [5-16], which can be equally interpreted as a saturable absorption effect [1-4, 17-20], the observed phase shift arising from the temporal absorption asymmetry [27] associated with the finite relaxation time (Fig 14). Indeed, the theoretical expressions for the modulation phase are mathematically identical [20], implying that CPO experiments alone cannot distinguish between slow light and saturable absorption. In this respect they bear a close resemblance to slow light experiments utilising the z-scan technique [28].

The analysis of wave interactions in saturable absorbers shows that the perturbation terms for hole-burning and absorption saturation are of equal magnitude, which supports this conclusion, but the latter is independent of frequency, allowing for broadband saturation of the absorption without hole-burning [22]. Saturable absorption theory, like CPO theory, predicts the appearance of the narrow hole of width $\sim 1/T_1$ observed in the modulation frequency spectrum and power broadening at higher incident intensities. Since virtually all CPO experiments have been performed with broadband spectral sources on time scales comparable with the population relaxation time $T_1$ in media where this is large compared with the phase coherence time $T_2$, it is appropriate to use a rate equation analysis [29]. However, the effect of doing so is to replace the coherent interaction of an amplitude modulated wave of frequency $\Omega_0$ (with sidebands at $\Omega_0 \pm \omega_0$) with the intensity dependent incoherent interaction of a broadband source, which is incapable of producing the narrow coherent hole in the optical spectrum required for slow light propagation [22]. Indeed, we have seen that slow light effects in alexandrite, ruby, $Er^{3+}:Y_2SiO_5$, bacteriorhodopsin, semiconductor waveguides and erbium doped optical fibres can be quantitatively evaluated in terms of saturable absorption theory. Similarly, gain saturation has been suggested for interpreting fast light effects in fibre optic amplifiers [16] and quantum dots [4]. Only where phase shifts and pulse delays exceeding the limits set by absorption/gain saturation are observed may more credence be accorded CPO experiments, as in recent experiments on semiconductor optical amplifiers [30, 31].

Separation of the pump and probe beams would seem a minimum requirement for distinguishing between hole burning and saturable absorption in slow light experiments e.g. employing a transverse pump and axial probe, enabling controlled bleaching of the homogeneously broadened absorption line without creating a coherent hole [32], or utilising orthogonally polarised pump and probe beams, thereby reducing the depth of the coherent hole by a factor of three [22], which should produce a significant reduction in the slow light effect. Stepanov and Hernández have recently observed controlled phase shifts in erbium fibre using widely separated pump and probe wavelengths, indicating saturation of the absorption band as a whole, rather than hole-burning within it [26]. The creation of a coherent hole in the absorption line can only be observed when a sufficiently narrow line source, such as a tunable diode laser, is used to independently scan the absorption spectrum within the homogeneous linewidth [12]. This is not the case for the single beam CPO experiments discussed above, which may be more correctly described as intensity modulation experiments for determining the frequency response of the non-linear medium [1]. These considerations do not conflict with the interpretation of amplitude modulation spectroscopy using a broad-band source, which is adequately described by saturable absorption theory [17-20], but do bring into question the observation of slow light with this technique. The fallacy lies in the interpretation of the observed phase shift as a signal transit time, which is then used to derive an apparent group velocity $v_g = L/\tau_d \geq L/\tau_s$, which can be made arbitrarily small (and the group index arbitrarily large) for an indefinitely thin sample ($L \Rightarrow 0$) of given optical density $\alpha_0 L$. A strong indication of the true interpretation is that in most cases the observed delay $\tau_d$ never exceeds the metastable lifetime $\tau_s$ and more often is just a fraction of it, whereas the transit time for slow light clearly depends on sample length and should exceed the relaxation time of the non-linear absorber in a sufficiently long sample [33].

5. Conclusion

The results of slow light experiments based on transmission of intensity modulated light by saturable absorbers are consistent with saturable absorption theory and arise from phase shifts caused by the finite relaxation time of the intensity dependent absorption [17-20]. As such they do not provide a clear demonstration of group velocity reduction. It is only when these facts are taken into account that more credible CPO experiments on slow light can be designed and carried out [12].

*Acknowledgement*

*My thanks to Valerii Zapasskii for bringing this problem to my attention, which entailed revisiting a 40 year-old field of theoretical research. Thanks also to Mike van der Poel of the Nanophotonics Group at DTU Lyngby for providing publications on pulse propagation in quantum devices, and to Professor Bruno Macke of the Université des Sciences et Technologies de Lille for kindly supplying e-prints of his work with Bernard Ségard and for sharing his views on 'slow light'.*

*Appendix*

*Equivalence of expressions for the modulation phase*

The dependence of the pump intensity ratio β on depth z in a saturable absorber is given by [5]

$$\frac{\partial \beta}{\partial z} = -\frac{\alpha_0 \beta}{1+\beta} \qquad (A1)$$

where $\alpha_0$ is the unsaturated absorption coefficient. The modulation phase dependence is [20]

$$\frac{\partial \varphi}{\partial z} = \frac{\alpha_0 \beta}{1+\beta} \frac{\omega_m \tau_s}{(1+\beta)^2 + (\omega_m \tau_s)^2} \qquad (A2)$$

substituting eq (A1) in eq (A2) and re-arranging

$$\frac{\partial \varphi}{\partial \beta} = -\frac{\omega_m \tau_s}{(1+\beta)^2 + (\omega_m \tau_s)^2} \qquad (A3)$$

hence
$$\varphi = \varphi_0 - \arctan \frac{1+\beta}{\omega_m \tau_s} \qquad (A4)$$

Evaluating eq (A4) over the interaction length L yields the overall modulation phase change

$$\Delta \varphi = -\arctan \frac{\omega_m \tau_s (\beta(0) - \beta(L))}{(1+\beta(0))(1+\beta(L)) + (\omega_m \tau_s)^2} \qquad (A5)$$

With the substitution $\beta(L) = \beta(0)T_s$ we obtain the expression for the modulation phase defined by eqns (11) and (12) in [20], which is identical to the saturable absorption phase shift in Sec 2 eq (3) above.

*Approximate formula for the time delay*

For dynamic equilibrium     $\ln(T_s/T_0) = \beta(1-T_s)$         (A6a)

With                         $\ln T_0 = -\alpha_0 L$              (A6c)

and                          $\ln T_s \approx -(1-T_s)$ for $1-T_s \ll 1$   (A6b)

we have                      $1-T_s \approx \alpha_0 L/(1+\beta)$  (A6d)

Substituting eq (A6d) in eq (A5) and evaluating for $\varphi \ll 1$ we find

$$\tau_d \approx \tau_s \frac{\beta}{1+\beta} \frac{\alpha_0 L}{(1+\beta)^2 + (\omega_m \tau_s)^2} \qquad (T_s \approx 1) \qquad (A7)$$

identical with the expression used to analyse slow light experiments when $\tau_d \ll \tau_s$ [6-9]

References

1. V S Zapasskii G G Kozlov Opt Spectrosc **100** 419-424 (2006)

2. K W Smith L Allen Opt Commun **8** 166-170 (1973)

3. M Adachi Y Masumoto Phys Rev B **40** 2908 (1989)

4. M van der Poel J Mørk J M Hvam Opt Express **13** 8032 (2005)

5. M S Bigelow N N Lepeshkin R W Boyd Phys Rev Lett **90** 113903 (2003)

6. E Baldit K Bencheikh P Monnier J A Levenson V Rouget Phys Rev Lett **95** 143601 (2005)

7. P Wu D V G L N Rao Phys Rev Lett **95** 253601 (2005)

8. C S Yelleswarapu R Philip F J Aranda B R Kimball D V G L N Rao Opt Lett **32** 1788-1790 (2007)

9. C S Yelleswarapu S Laoui R Philip D V G L N Rao Opt Express **16** 3844-3852 (2008)

10. S Melle O G Calderón F Carreño E Cabrera M A Antón S Jarabo Opt Commun **279** 53-63 (2007)

11. P Palinginis F Sedgwick S Crankshaw M Moewe C J Chang-Hasnain Opt Express **13** 9909 (2005)

12. P C Ku F Sedgwick C J Chang-Hasnain et al. Opt Lett **29** 2291 (2004)

13. L W Hillman R W Boyd J Krasinski C R Stroud Opt Commun **45** 416 (1983)

14. M S Malcuit R W Boyd L W Hillman J Krasinski C R Stroud Jr J Opt Soc Am B **1** 73 (1984)

15. M S Bigelow N N Lepeshkin R W Boyd Science **301** 200 (2003)

16. A Schweinsberg N N Lepeshkin M S Bigelow R W Boyd S Jarabo Europhys Lett **73** 218 (2006)

17. A C Selden Electron Lett **7** 287 (1971); Brit J Appl Phys **18** 743 (1967), J Phys D **3** 1935 (1970)

18. M A Kramer R W Boyd L W Hillman C R Stroud Jr J Opt Soc Am B **2** 1444 (1985)

19. B Macke B Ségard Phys Rev A **78** 013817 (2008)

20. J Mørk R Kjær M van der Poel K Yvind Opt Express **13** 8136 (2005)

21. M S Bigelow N N Lepeshkin R W Boyd J Cond Matt Phys **16** 1321 (2004)

22. S E Schwarz T Y Tan Appl Phys Lett **10** 4-7 (1967)

23. G S Agarwal T N Dey Phys Rev A **73** 043809 (2006)

24. F Öhman K Yvind J Mørk Opt Express **14** 9955-9962 (2006)

25. S Melle O G Calderón C E Caro E Cabrera-Granado M A Antón F Carreño Opt Lett **33** 827 (2008)

26. S Stepanov E H Hernández Opt Lett **33** 2242 (2008)

27. H Wang Y Zhang H Tian N Wang L Ma P Yuan Appl Phys B **92** 487-491 (2008)

28. I Guedes L Misoguti S C Zilio Opt Express **14** 6201 (2006)

29. G Piredda R W Boyd J Eur Opt Soc **2** 07004 (2007)

30. W Xue Y Chen F Öhman S Sales J Mørk Opt Lett **33** 1084 (2008)

31.  B Pesala F Sedgwick A Uskov C J Chang-Hasnain J Opt Soc Am B **25** C46 (2008)
32. R N Shakhmuratov A Rebane P Mégret J Odeurs Phys Rev A **71** 053811 (2005)

33. R W Boyd D J Gauthier A L Gaeta A E Willner Phys Rev A **71** 023801 (2005)

*A C Selden*                                                                    *December 2008*

Figure captions

Fig. 1 'Hole-burning' and 'power-broadening' in ruby pumped by a frequency modulated argon ion laser. Saturable absorption curves fitted to the experimental data of [13].

Fig. 2 Time delay $\tau_d$ and relative modulation attenuation $A(\omega)$ of the transmitted intensity vs. modulation frequency for pump powers $P_0 = 0.1W$, $0.25W$ incident on a ruby sample. The curves are derived from saturable absorption theory, the plotted points are the experimental data for 'slow light' [5]. Fitting parameters: $\beta = 1.5$ for $P_0 = 0.25W$, $\beta = 0.6$ for $P_0 = 0.1W$.

Fig. 3 Modulation frequency spectra for alexandrite [15]. Saturable absorption theory (curves) fitted to the experimental data of [15] with $\tau_s = 260$ μs (anti-hole) and $\tau_s' = 20$ ms (hole).

Fig. 4 Comparison of saturable absorption theory with experimental data for time delay $\tau_d$ vs. incident power $P_0$ for CPO in $Er^{3+}$:$Y_2SiO_5$ at 0, 10, 40 and 80 Hz modulation frequencies [6]. The dashed curves represent inhomogeneous broadening [23].

Fig. 5 Signal delay $\tau_d$ in bacteriorhodopsin (bR) thin film vs. intensity $I_b$ of blue light. The points are the experimental data of Wu and Rao [7], the saturable absorption curve (-----) is derived from eq (3), with intensity dependent lifetime $\tau_s(I_b) = \tau_s/(1+AI_b)$, where $\tau_s = 0.28$ s and $A = 1.18$ mW$^{-1}$. The full curve shows the fit of the 'slow light' formula (eq (1) of [7]) to the data.

Fig. 6 Signal delay $\tau_d$ vs. incident power $P_0$ for bR aq. solution with fitting parameters $P_s = 6$ mW and $\tau_s = 60$ ms vs. the values $P_s = 21$ mW and $\tau_s = 77$ ms reported in [8]. These differences may result from averaging rather than integrating over the optical path.

Fig. 7 Correlation of signal delay $\tau_d$ and attenuation $A(\omega)$ vs. modulation frequency for bR aq. solution. Saturable absorption curves (— $A(\omega)$, --- $\tau_d$) fitted to the experimental data of [9].

Fig. 8 Reverse saturable absorption theory (curve) provides a good fit to the experimental data for 'fast light' in bR aq. solution [9] with $\tau_s = 30$ ms.

Fig. 9 Saturable absorption curves matched to the GaAs QW waveguide results of Palinginis et al for the parameters $P_s = 123$ mW and $\tau_s = 5.9$ ns derived from power broadening [11].

Fig. 10 Saturable absorption theory (curve) compared with the experimental data for group velocity reduction in a MQW structure [12], showing that saturable absorption agrees with CPO even when there is independent evidence of coherent hole burning within the heavy hole (H-H) absorption band.

Fig. 11 Erbium doped optical fibre results for the Er20 sample (20 dB/m) [10]. The fractional delay $F_d$ vs. modulation frequency curve for saturable absorption is an excellent fit to the experimental points for $P_0 = 1$ mW. Fitting parameters $P_s = 0.4$ mW, $\tau_s = 10.5$ ms [10].

Fig. 12 Maximum fractional delay $F_{max}$ vs. incident power $P_0$ for Er20 fibre [10]. Parameters $P_s$, $\tau_s$ as in Fig 11.

Fig. 13 Optimum modulation frequency $f_{opt}$ for maximum fractional delay vs. incident power $P_0$ for the Er20 fibre, showing the near-linear dependence of $f_{opt}$ on $P_0$ [10]. $P_s$, $\tau_s$ as in Figs 11, 12.

Fig. 14 Incident (upper) and transmitted (lower) signals calculated for ruby, showing signal delay corresponding to asymmetric absorption [27] and 'instantaneous' transmission of discontinuities.

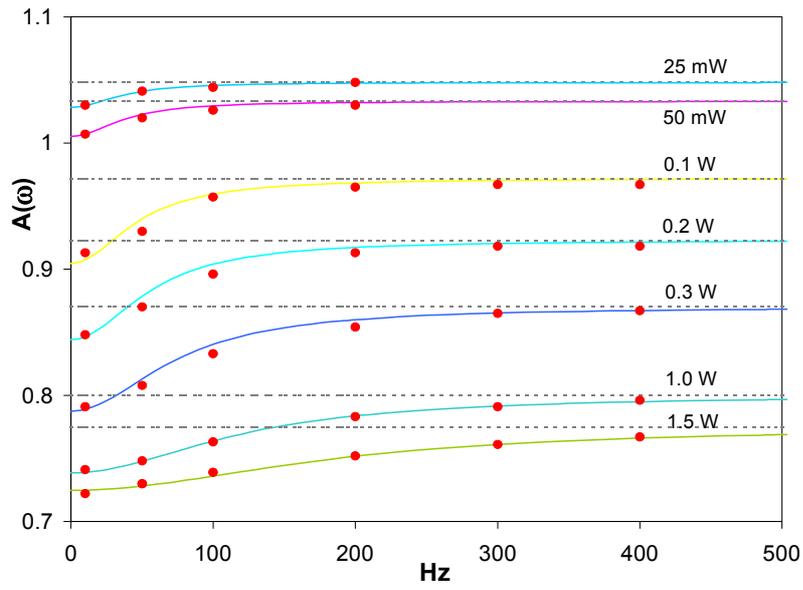

Fig 1

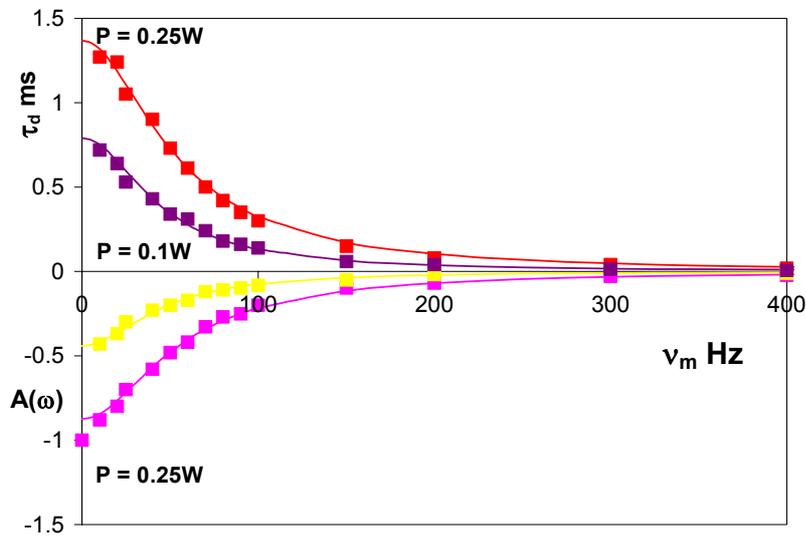

Fig 2

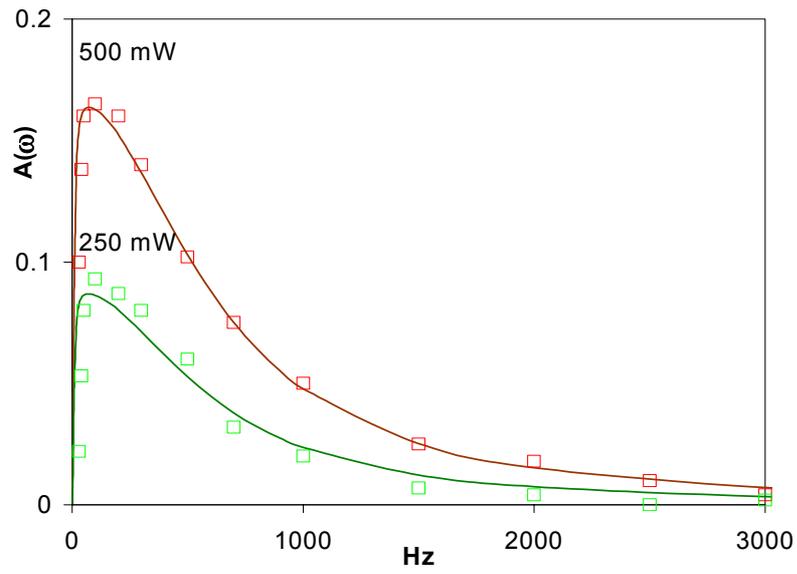

Fig 3

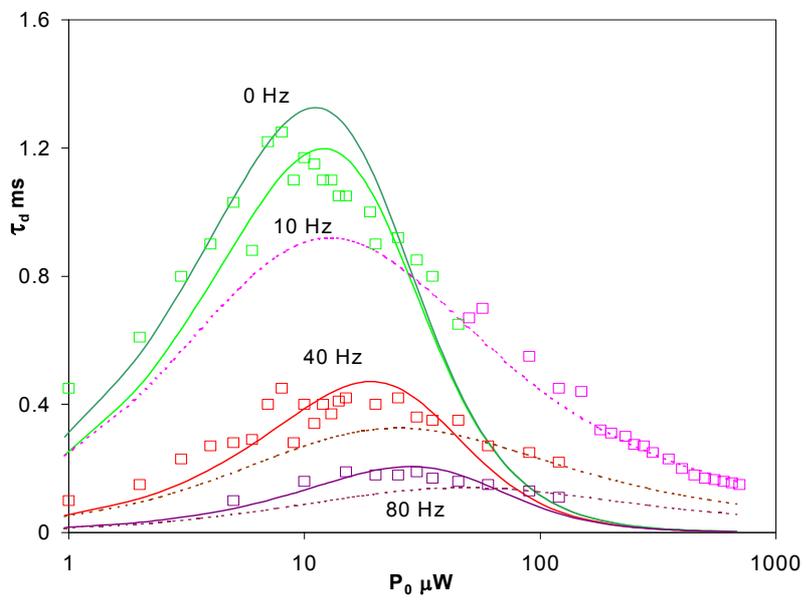

Fig 4

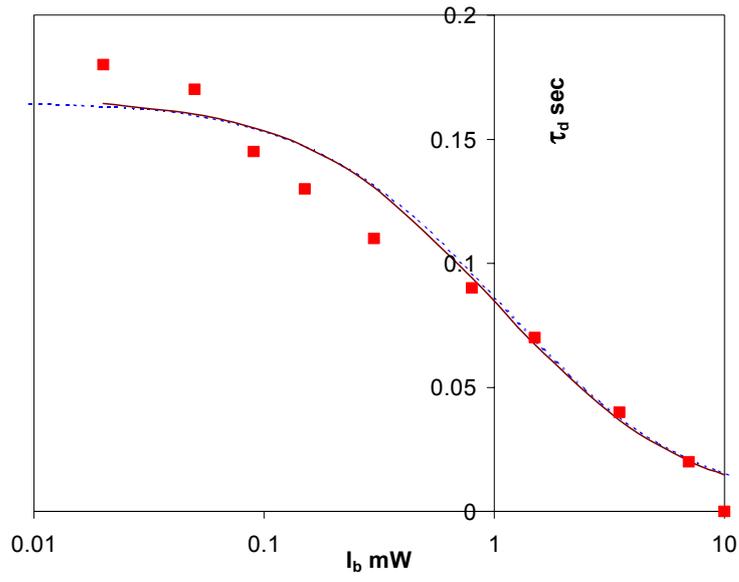

Fig 5

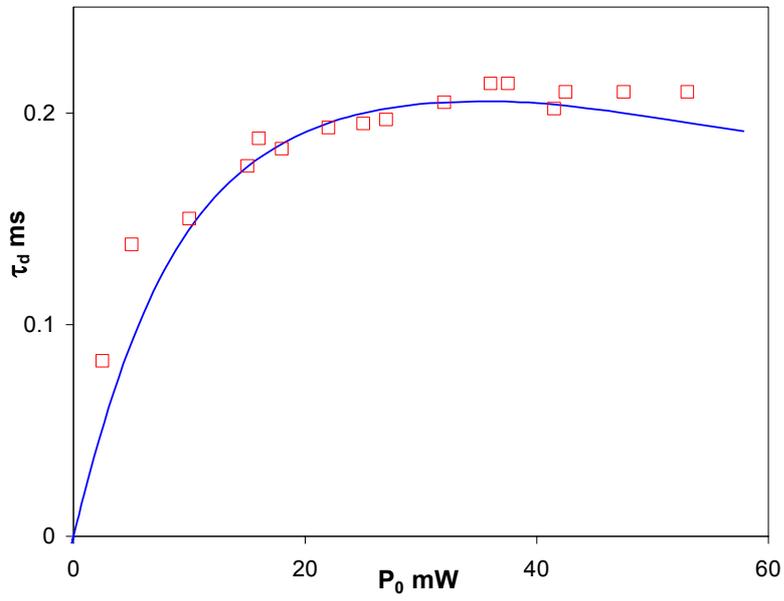

Fig 6

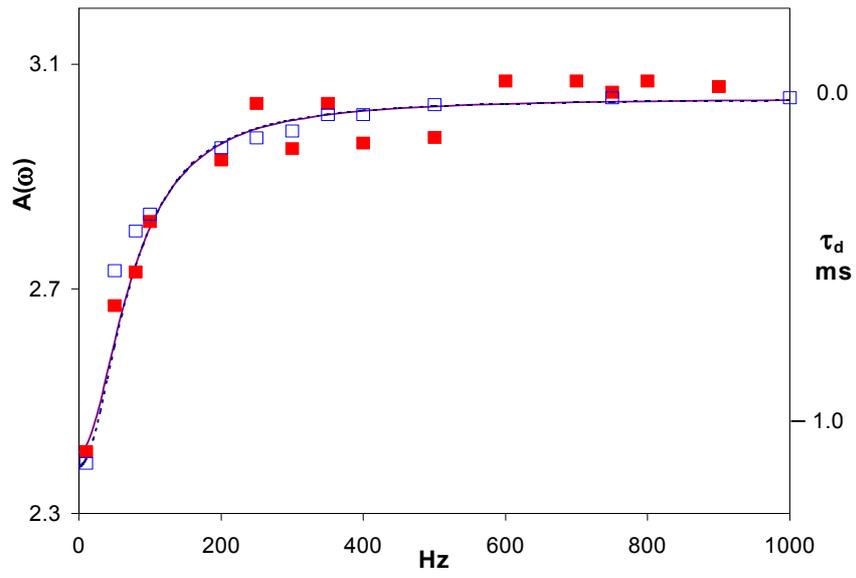

Fig 7

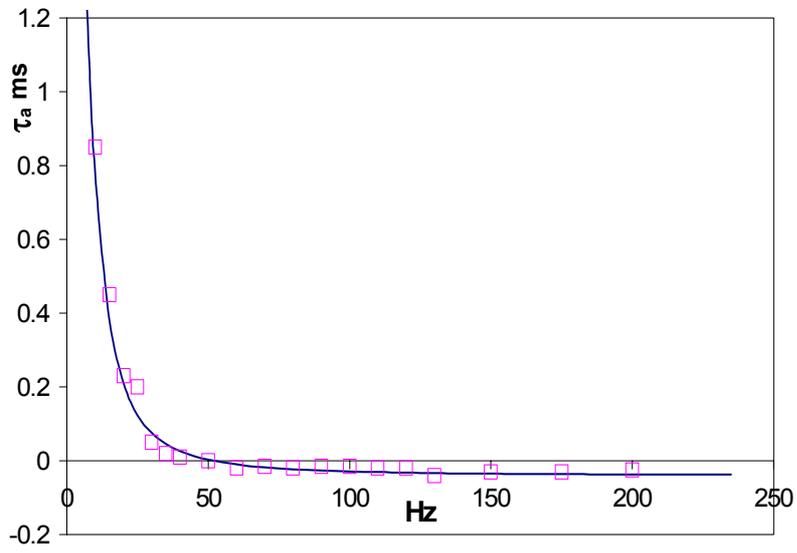

Fig 8

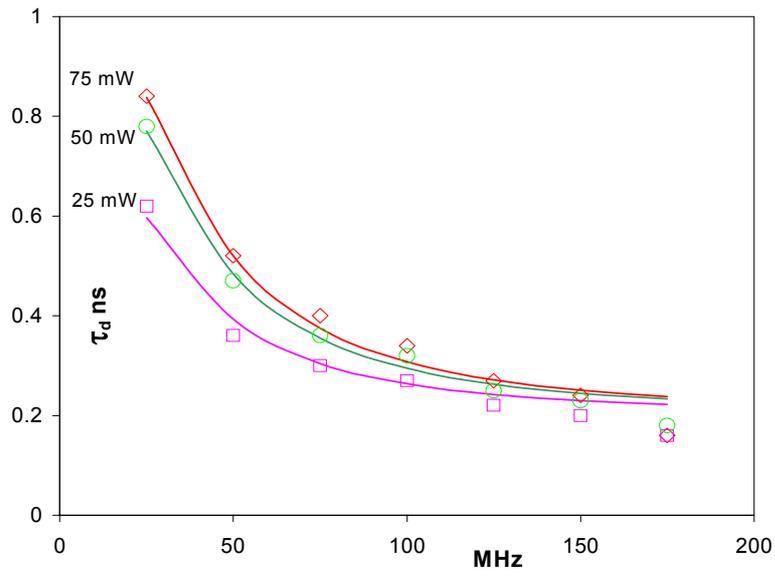

Fig 9

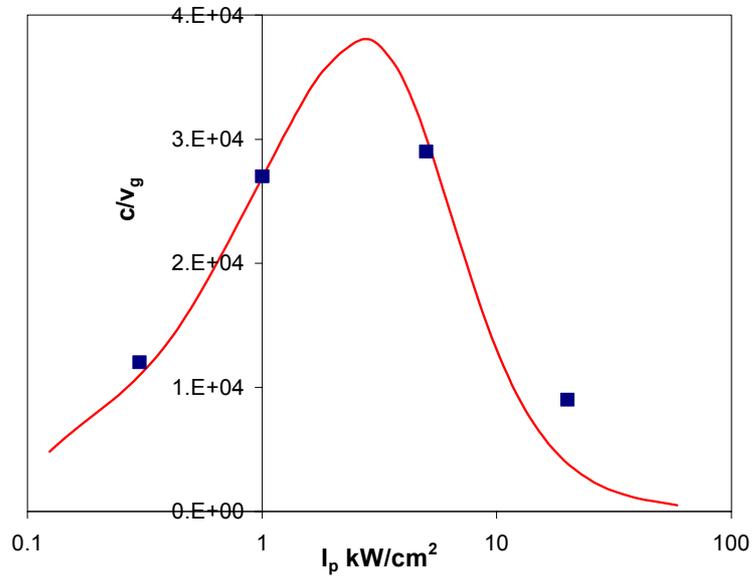

Fig 10

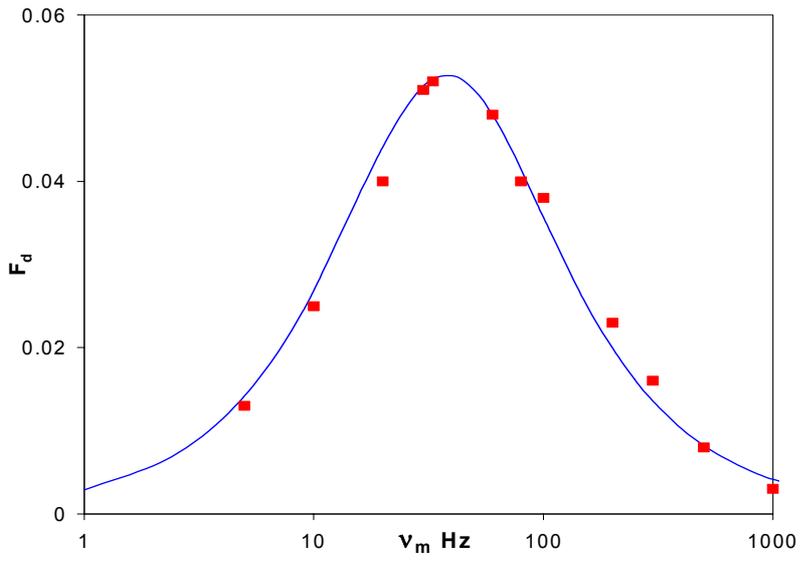

Fig 11

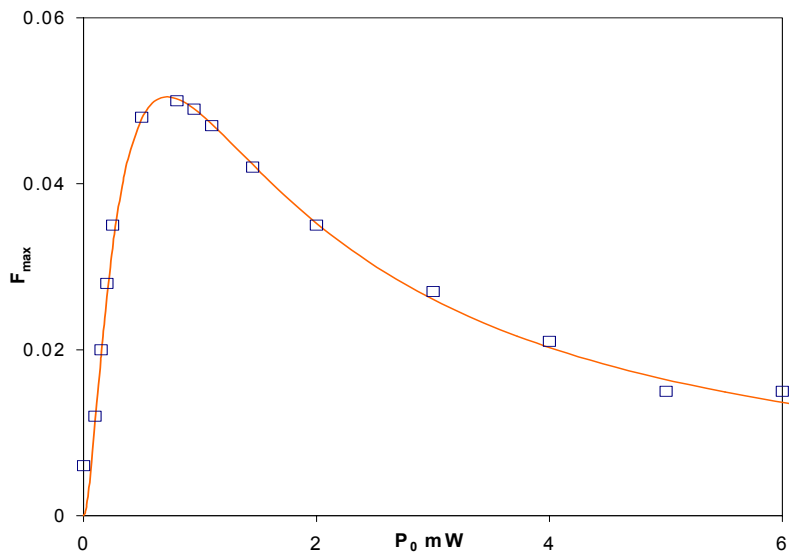

Fig 12

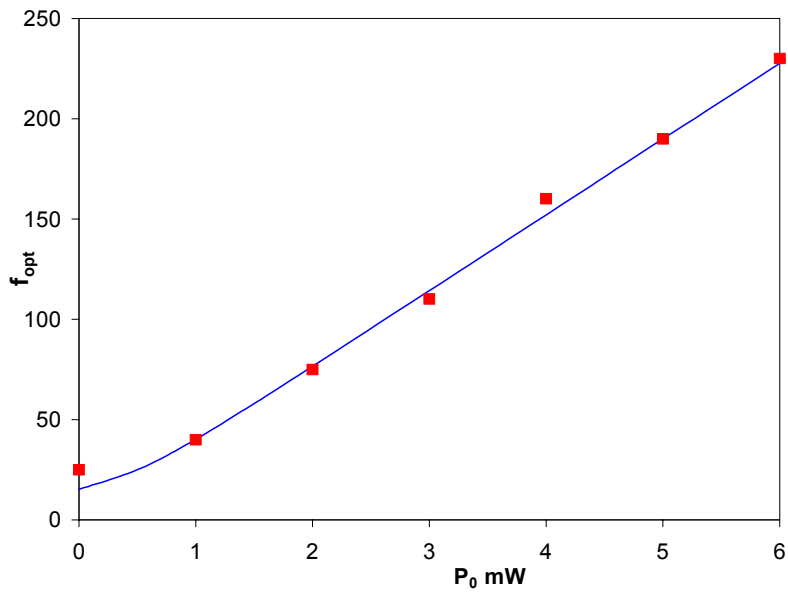

Fig 13

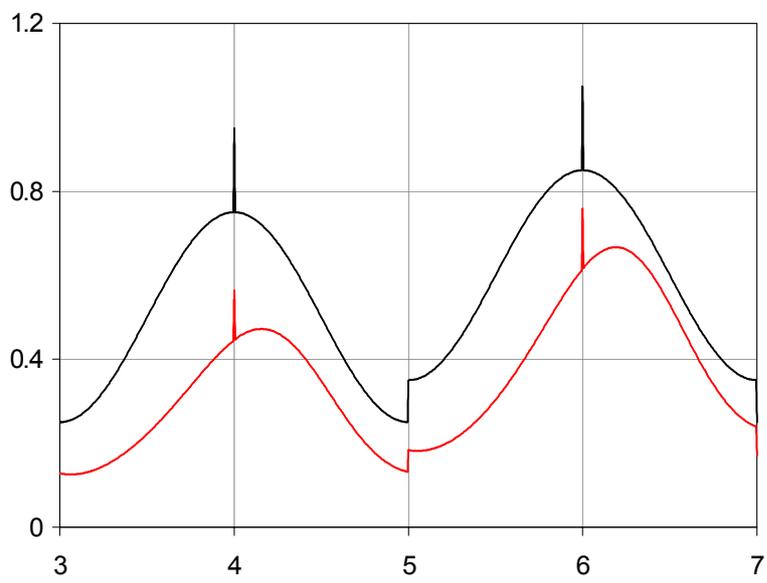

Fig 14